\def\year{2016}\relax
\documentclass[letterpaper]{article}
\usepackage{aaai16}
\usepackage{times}
\usepackage{helvet}
\usepackage{courier}
\usepackage{graphicx}
\usepackage{amsmath}
\title{Predictability of Popularity: Gaps between Prediction and Understanding}
\author{Benjamin Shulman \\ Dept. of Computer Science \\ Cornell University \\ bgs53@cornell.edu
   \And
   Amit Sharma\\ Microsoft Research \\ New York, NY \\ amshar@microsoft.com
   \And
   Dan Cosley \\ Information Science \\ Cornell University \\ drc44@cornell.edu
 }

\begin{document}

 \maketitle

%


%
%
%
%

\begin{abstract}
Can we predict the future popularity of a song, movie or tweet? Recent work suggests that although it may be hard to predict an item's popularity when it is first introduced, peeking into its early adopters and properties of their social network makes the problem easier.
We test the robustness of such claims by using data from social networks spanning music, books, photos, and URLs. We find a stronger result: not only do predictive models with peeking achieve high accuracy on all datasets, they also generalize well, so much so that models trained on any one dataset perform with comparable accuracy on items from other datasets.

Though practically useful, our models (and those in other work) are intellectually unsatisfying because common formulations of the problem, which involve peeking at the first small-$k$ adopters and predicting whether items end up in the top half of popular items, are both too sensitive to the speed of early adoption and too easy.
Most of the predictive power comes from looking at how quickly items reach their first few adopters, while for other features of early adopters and their networks, even the direction of correlation with popularity is not consistent across domains.  Problem formulations that examine items that reach $k$ adopters in about the same amount of time reduce the importance of temporal features, but also overall accuracy, highlighting that we understand little about why items become popular while providing a context in which we might build that understanding.
\end{abstract}




\noindent How does a book, song, or a movie become popular? The question of how cultural artifacts spread through social networks has captured the imagination of scholars for decades. Many factors are cited as important for an item to spread \textit{virally} through social networks and become popular: its intrinsic quality \cite{gladwell2006-formula,simonoff2000},  
 the characteristics of its initial adopters \cite{gladwell2006}, the emotional response it elicits \cite{berger2012}, and so on. Often, explanations are used to justify the popularity of different items after the fact \cite{berger2013book}, making it hard to apply these explanations to new events \cite{watts2011}.

Online social networks allow us to observe individual-level traces of how items are transferred between people, allowing more precise modeling of the phenomenon. Predicting the future popularity of an item based on attributes of the item and the person who introduced it has emerged as a useful problem, both to understand processes of information diffusion and to inform content creation and feed design on social media platforms.
For example, Twitter's managers may want to highlight new tweets that are more likely to become popular, while its users may want to learn from characteristics of popular tweets to imporve their own.

In general, even with detailed information about an item's content or the person sharing it, it is hard to predict which items will become more popular than others \cite{bakshy2011,martin2016}.  The problem becomes more tractable when we are allowed to \textit{peek} into the initial spread of an item. The intuition is that early activity data about the speed of adoption, characteristics of people who adopt it and the connections between them might predict the item's fate. This intuition shows encouraging results for both predicting the final popularity of an item \cite{szabo2010,pinto2013,zhao2015} and whether an item will end up in the top 50\% of popular items \cite{cheng2014,romero2013,weng2013}.

Buoyed by these successes, one might conclude that the availability of rich features about the item and social network of early adopters has helped us understand why items become popular.  However, past work studies individual datasets and varying versions of the prediction problem, making it hard to compare results.  For instance, studies disagree on the direction of the effect of network structural features on item popularity \cite{lerman2010,romero2013}.

In this paper, we try to unify these observations on popularity prediction through studying different problem formulations and kinds of features over a wide range of online social networks.  Using an existing formulation that predicts whether the final popularity of items is above the median based on features of the first five adopters \cite{cheng2014}, we confirm past work \cite{szabo2010} showing that features about those adopters and their social network are at best weak predictors of popularity compared to temporal features. For instance, a single temporal heuristic---the average rate of early adoption---is a better predictor than all non-temporal features combined across all four websites.  Further, models trained on one dataset and tested on others using temporal features generalize fairly well, while those that use network structural features generalize badly.

In one reading, this is a useful contribution: peeking-based popularity models that include temporal information achieve up to 83\% accuracy on Twitter and generalize well across datasets.  From a practical standpoint, we encourage content distributors to use temporal features for predicting the future success of items.

Intellectually, however, our finding is not very satisfying.  Rather than identifying features that shed light on why items become popular, we mostly see that items that become popular fast are more likely to achieve higher popularity in the end. Rapid adoption may be a signal of quality, interestingness, and eventual popularity---but doesn't tell us why. The effect might also be driven by cumulative advantage \cite{frank2010winner,watts2011}: items that receive attention early have more chances to spread through via interfaces that highlight popular or trending items.

An alternative formulation of the problem that reduces the effect of temporal features lets us see just what early adopter and network features tell us.  This formulation, called Temporal Matching, compares items that achieve similar levels of popularity in the same amount of time, rather than the more common formulation of looking at the first $k$ adopters regardless of the time it takes to reach $k$.  Controlling for the average rate of an item's adoption turns  popularity prediction into a hard problem. Using the same features as before, prediction accuracy across all datasets drops below 65\%.
Such a decrease in accuracy underscores the importance of choosing problem formulations that highlighting relevant phenomena in popularity evolution.  Current models may fare well on certain formulations, but there is still much to learn about how items become popular.

\begin{table*}[t]
\center
\begin{tabular}{llcccc}
\textbf{Study} &  \textbf{Problem Formulation} & \textbf{Content} & \textbf{Structural} & \textbf{Early Adopters} & \textbf{Temporal} \\

Bakshy et al. (2011) & Regression (no peeking) & n & -- & \textbf{Y} & -- \\
Martin et al. (2016) & Regression (no peeking) & n & -- & \textbf{Y} & -- \\

Szabo et al. (2010) & Regression & -- & n & -- & \textbf{Y} \\
Tsur et al. (2012)  & Regression & \textbf{Y} & \textbf{Y} & -- & \textbf{Y} \\
Pinto et al. (2013) & Regression & -- & -- & -- & \textbf{Y} \\
Yu et al. (2015) & Regression & -- & n & -- & \textbf{Y} \\
Romero et al. (2013) & Classification \small{($k=\{1000,2000\}, n=50\%$)} & -- & \textbf{Y} & -- & -- \\
Cheng et al. (2014)  & Classification \small{($k=5,n=50\%$)} & n & \textbf{Y} & \textbf{Y} & \textbf{Y} \\
Lerman et al. (2008)  & Classification \small{($k=10, n=80\%$)} &  -- & \textbf{Y} & \textbf{Y} & -- \\
Weng et al. (2013) & Classification \small{($k=50, n=\{70,80,90\%\}$)} & -- & \textbf{Y} & n & -- \\

\end{tabular}
\caption{A taxonomy of problem formulations for popularity prediction, along with importance of feature categories.  \textbf{Y} means that the features in the category were useful for prediction, n means they were tried but not as useful, and -- that they were not studied. Most studies report temporal and structural features as important predictors. }
\label{tab:past-features}
\end{table*}

\section{Formulations of the prediction problem}
\label{sec:problem-formulation}
We start by identifying two key dimensions to consider when defining the popularity prediction task: how much peeking into early activity on an item is allowed, and whether the task is a regression or classification.  For ease of exposition, we use \textit{item} to denote entities that are consumed in online social networks. \textit{Adoption} refers to an explicit action or endorsement of an item, such as loving a song, favoriting a photo, rating a book highly or retweeting a URL.  Finally, we define \textit{popularity} of an item as the number of people who have adopted it.

\subsection{Predicting apriori versus peeking into early activity}
Predicting popularity \textit{a priori} for items such as movies \cite{simonoff2000} or songs \cite{pachet2012} has long been considered a hard problem. One of the most successful approaches has been to gauge audiences' interest in an item before it is officially released, such as by measuring the volume of tweets \cite{asur2010} or search queries \cite{goel2010}.  Such methods can work well for mainstream, popular items for which there might be measurable prior buzz, but are unlikely to be useful for genuinely new items such as tweets or photos uploaded by users.

For such items, popularity prediction is tricky, even when precise data about the content of each tweet and the seed user's social network is known. On Twitter, models with extensive content features such as the type of content, its source and topic, crowdsourced scores of interestingness, and features about the seed user such as indegree and past popularity of tweets are only able to explain less than half of the variance in popularity \cite{martin2016}.
Further, the content features are usually less important than features of the seed user  \cite{bakshy2011,martin2016,jenders2013}.

In response, scholars have suggested modified versions of the problem where one peeks into early adoption activity for an item.
In studies on networks including Facebook \cite{cheng2014}, Twitter \cite{lerman2010,zhao2015,tsur2012,kupavskii2013}, Weibo \cite{yu2015}, Digg \cite{lerman2010,szabo2010} and Youtube \cite{pinto2013}, early activity data consistently predicts future popularity with reasonable accuracy.  
In light of these results, we focus on the peeking variant of the problem in this paper.

\subsection{Classification versus regression}
In addition to how much data we look at, we must also specify what to predict.  A number of studies have used regression formulations, predicting an item's exact final popularity: the number of retweets for a URL \cite{bakshy2011}, votes on a Digg post \cite{lerman2010} or page views of a Youtube video \cite{szabo2010}.
However, we may often be more interested in popularity relative to other items rather than an exact estimate.  For example, both marketers and platform owners may want to select `up and coming' items to feature in the interface versus others\footnote{Such featuring makes some items more salient than others and surely affects the final popularity of both featured and non-featured items; typically, formulations of the problem look at very small slices of early activity, which presumably minimizes these effects.}.

These motivations lead nicely to a classification problem where the goal is to predict whether an item will be more popular then a certain percentage of other items.
For instance, Romero et al. predict whether the number of adopters of a hashtag on Twitter will double, given a set of hashtags with the same number of initial adopters  \cite{romero2013}. Cheng et al. generalize this formulation to show that predicting whether an item will double its popularity is equivalent to classifying whether an item  becomes more popular than the median and study this question in the case of Facebook photos that received at least five adopters \cite{cheng2014}.  Besides the practical appeal of classifying popular items, classification is also a simpler task than predicting the actual number of adoptions \cite{bandari2012}, thus providing a favorable scenario for evaluating the limits of predictability of popularity.
Therefore, we focus on the classification problem in this paper.

\subsection{Our problem: Peeking-based classification}
Based on the above discussion, the general peeking-based classification problem can be stated as:

\begin{quote}
\noindent \textbf{P1: }\textit{Given a set of items and data about their early adoptions, which among them are more likely to become popular?}
\end{quote}

This question has a broad range of formulations based on how we define the early activity period, how much activity we are allowed to poke at, and how we define \textit{popular}.  The early activity period may be defined in terms of time elapsed $t$ since an item's introduction \cite{szabo2010}, or in terms of a fixed number $k$ of early adoptions \cite{romero2013}. Fixing the early activity period in terms of number of adoptions has the useful side-effect of filtering out items with less than $k$ adoptions overall, both making the problem harder and eliminating unpopular (thus often uninteresting) items.  For this reason, most past work on peeking-based classification defines early activity in terms of the number of adoptions $k$.

The popularity threshold for what is ``popular'' may also be set at different percentiles ($n\%$). Table~\ref{tab:past-features} summarizes past work based on their choices of problem formulation and choice of $(k, n)$.  One common approach is to collect all items that have $k$ or more adoptions, then peek into the first $k$ adoptions and predict whether eventual popularity of items lies above or below the median \cite{cheng2014}. We call this Balanced Classification since there are guaranteed to be an equal number of high and low popularity items. Another variation is to only consider the top-\textit{n} percentile of items as high popularity \cite{lerman2008}, a formulation that is arguably better-aligned with most use cases around content promotion than Balanced Classification.  However, it is also harder than Balanced Classification; for this reason, and to continue to align with prior work, we focus on Balanced Classification.

While restricting to items with $k$ adoptions helps to level the playing field because it provides a set of comparably popular items to study, it ignores the \textit{time taken} to reach $k$ adoptions.  Based on prior work, our suspicion is that in this formulation temporal features dominate the others.  To control for this temporal signal, we later introduce a problem formulation where both $k$ and $t$ are fixed. That is, we collect all items that received exactly $k$ adoptions in a given time period $t$, and then predict which of them would be in the top half of popular items. We call this the Temporally Matched Balanced Classification problem, and as we will see, changing the definition has a profound impact on the quality of the models.

\section{Choosing features}
\label{sec:features}
We now turn to the selection of features for prediction.
Part of the allure of modeling is that the features that prove important might give information about \textit{why} some items become popular in ways that could be both practically and scientifically interesting.
Features used in prior work can be broadly grouped into four main categories:
content, structural, early adopters and temporal \cite{cheng2014}.  Table~\ref{tab:past-features} shows which feature categories were used in prior studies, with cells in bold representing features that were reported to be useful for prediction. While all feature categories have been reported to be important contributors to prediction accuracy in at least some studies, temporal and structural features are frequently reported as important.

Temporal patterns of early adoption---how quickly the early adopters act---are a major predictor of popularity.  Szabo and Huberman show that temporal features alone can predict future popularity reliably \cite{szabo2010}.  When information about the social network or its users is hard to obtain, utilizing temporal features can be fruitful, achieving error rates as low as 15\% in a regression formulation \cite{pinto2013,zhao2015}.
A natural next question is to ask how much these errors can be decreased by adding other features when we do have such information.

Features about the seed user and early resharers---collectively called early adopters---also matter. On Twitter, for example, the number of followers of the seed user and the fraction of her past tweets that received retweets increase the accuracy of predictions \cite{tsur2012}.  Information about other early adopters is also useful for predicting photo cascades in Facebook \cite{cheng2014}.

The structure of the underlying social network also has predictive power \cite{lerman2008,romero2013,cheng2014}. However, these studies do not agree on the direction of effect of these features. For instance, on Digg, low network density is connected with high popularity \cite{lerman2008}, but on Twitter, both very low and very high densities are positively correlated with popularity \cite{romero2013}. Their intuition is that a lower network density indicates that the item is capable of appealing to a general audience, while a higher network density indicates a tight-knit community supporting the item, both of which can be powerful drivers for an item's popularity.

Finally, while Tsur et al. report content features to be useful \cite{tsur2012}, most studies find content features to have little predictive power (Table~\ref{tab:past-features}).  Even for domains such as songs or movies where item information is readily available, content features are not significantly associated with item popularity \cite{pachet2012}. Further, content features do not generalize well; it is hard to compute generalizable content features across different item domains. For these reasons, we do not consider content features in this work.
 
\subsection{Features}
\label{sec:our-features}
Based on the above discussion, we use the following categories of features, with the aim of reproducing and extending the features used in past work \cite{cheng2014}: temporal, structural, and early adopters. To these we add a set of novel features based on preference similarity between early adopters.

\subsubsection{Temporal.}
\label{sec:temporal-features}
These features have to do with the speed of adoptions during the early adoption period between the first and $k$th adoption.  This leads to a set of features that focus on the rate of adoption:
\begin{itemize}
\item $time_i$: time between the initial adoption and the $i^{th}$ adoption ($2\leq i\leq k$). \cite{zhao2015,maity2015,weng2013}
\item $time_{1...k/2}$: Mean time between adoptions for the first half (rounded down) of the adoptions.
\item $time_{k/2...k}$: Mean time between adoptions for the last half (rounded up) of the adoptions.
\end{itemize}

\subsubsection{Structural.}
These features have to do with the structure of the network around early adopters and can be broken down into two sub-categories: ego network features that relate the early adopters to their local networks, and subgraph features that consider only connections between the early adopters.

\noindent \textit{Early adopters' ego network features}
  \begin{itemize}
  \item $in_{i}$: Indegree of the $i^{th}$ early adopter ($2 \leq i \leq k$). This is a proxy for the number of people who may be exposed to an early adopter's activity. For undirected networks, this will simply be the degree, or the number of friends of an early adopter. \cite{bakshy2011,zhao2015}
  \item $reach$: Number of nodes reachable in one step from the early adopters.
  \item $connections$: Number of edges from early adopters to the entire graph. \cite{romero2013}
  
  \end{itemize}

\noindent \textit{Early adopters' subgraph features}
  \begin{itemize}
  \item $indegree_{sub}$: Mean indegree (friends or followers) for each node in the subgraph of early adopters. \cite{lerman2008}
  \item $density_{sub}$: Number of edges in the subgraph of early adopters. \cite{romero2013}
  \item $cc_{sub}$: Number of connected components in the subgraph of early adopters. \cite{romero2013}
  \item $dist_{sub}$: Mean distance between connected nodes in the subgraph of early adopters. This is meant to measure how far the item has spread in the initial early adopters, similar to the cascade depth feature by Cheng et al.
  \item $sub\_in_{i}$: Indegree of the $i^{th}$ adopter on the subgraph ($1 \leq i \leq k$). \cite{lerman2008}
  \end{itemize}

\subsubsection{Features of early adopters.}

These features capture information about early adopters, such as their popularity, seniority, or activity level, which might be proxies for their influence. They can be divided into two sub-categories: features of the first user to adopt an item (root), and features averaged over other early adopters (resharers).

\noindent \textit{Root features}
\begin{itemize}
\item $activity_{root}$: Number of adoptions in the four weeks before the end of the early adoption period.  This is similar to a measure used by Cheng et al. which measured the number of days a user was active. \cite{cheng2014,petrovic2011,yang2010}
\item $age_{root}$: Length of time the user has been registered on the social network.
\item $popularity_{root}$: Number of friends or followers on the social network. \cite{lerman2008,tsur2012}
\end{itemize}

\noindent \textit{Resharer features}
\begin{itemize}
\item $activity_{resharer}$: Mean number of adoptions in the four weeks before the end of the early adoption period.
\item $age_{resharer}$: Mean length of time the users have been registered on the social network.
\item $popularity_{resharer}$: Mean number of friends or followers on the social network. \cite{tsur2012}
\end{itemize}

\begin{table*}[th]
\center
\begin{tabular}{lllll} \hline
\textbf{Dataset} & \textbf{Last.fm} & \textbf{Twitter} & \textbf{Flickr} & \textbf{Goodreads} \\ \hline
Number of users & 437k & 737k & 183k & 252k \\
Number of items & 5.8M & 64k & 10.9M & 1.3M \\
Number of adoptions & 44M & 2.7M & 33M & 28M \\
Mean adoptions & 7.6 & 41.8 & 3.0 & 21.4 \\
Median adoptions & 1 & 1 & 1 & 1 \\
Maximum adoptions & 11062 & 82507 & 2762 & 88027 \\ \hline
\end{tabular}
\caption{Descriptive statstics for users, items, and adoptions in each dataset. We use \textit{adoption} to mean loving a song on Last.fm, tweeting a URL on Twitter, favoriting a photo on Flickr, and rating a book on Goodreads. The average number of adoptions per item varies quite a bit, but the median popularity of 1 is consistent across datasets.}
\label{tab:descriptive}
\end{table*}

\subsubsection{Similarity}
To these previously tested features, we add features related to preference similarity between the early adopters.  As with network density, our intuition is that similarity between early adopters may matter in two ways: high similarity may signify a niche item, or one that people with similar interests are likely to adopt, while low similarity might indicate an item that could appeal to a wide variety of people.

Similarity was computed using the Jaccard index of two users' adoptions that occurred before the end of the early adoption period of the item in question.
We computed the median, mean and maximum of similarity between adopters because these give us an idea of the distribution of the affinity of the early adopters; we do not include users who had less than five adoptions before the item in question because they are likely to have little overlap.  The features we extracted are:
\begin{itemize}
\item $sim_{count}$: Number of similarities that could be computed between early adopters.
\item $sim_{mean}$: Mean similarity between early adopters.
\item $sim_{med}$: Median similarity between early adopters.
\item $sim_{max}$: Maximum similarity between early adopters.
\end{itemize}

\section{Data and Method}
\subsection{Datasets from four online social networks}
We build models using data from four different online social platforms: Last.fm, Flickr, Goodreads and Twitter.  These platforms span a broad range of online activity, including songs, photos, books and URLs; they also have a variety of user interfaces, use cases, and user populations.  These variations 
reduce the risk of overfitting to properties of a particular social network.

\begin{itemize}
\item
\textbf{Last.fm:} A music-focused social network where users can friend one another and love songs. We consider a dataset of 437k users and the songs they loved from their start date until February 2014 \cite{sharma2016}.

\item
\textbf{Flickr:} A photo sharing website where users can friend one another and favorite photos. We use data collected over 104 days in 2006 and 2007 \cite{cha2009}.

\item
\textbf{Goodreads:} A book rating website where users can friend one another and rate books. The dataset consists of 252k users and their ratings before August 2010.  Unlike the other sites, Goodreads users rate books; we consider any rating at or above 4 (out of 5) as an endorsement (adoption) of the book \cite{huang2012}.

\item
\textbf{Twitter:} A social networking site where users can form directed edges with one another and broadcast \textit{tweets}, messages no longer than 140 characters (as of 2010). The Twitter dataset consists of URLs tweeted by 737k users for three weeks of 2010 \cite{hodas2014}.
\end{itemize}

All of these websites have an active social network, providing an activity feed that allows users to explore, like, and reshare the items that their friends shared. The Last.fm feed shows songs that friends have to listened to or \textit{loved}, Flickr shows photos that friends have \textit{favorited}, Goodreads shows books that friends have \textit{rated}, and Twitter shows tweets with URLs that followees have \textit{favorited} or \textit{retweeted}. Thus, like past studies on online social networks such as Facebook, Twitter and Digg, we expect active peer influence processes that should make structural and early adopter features relevant.

\begin{figure}[t]
\centerline{\includegraphics[width=0.99\linewidth]{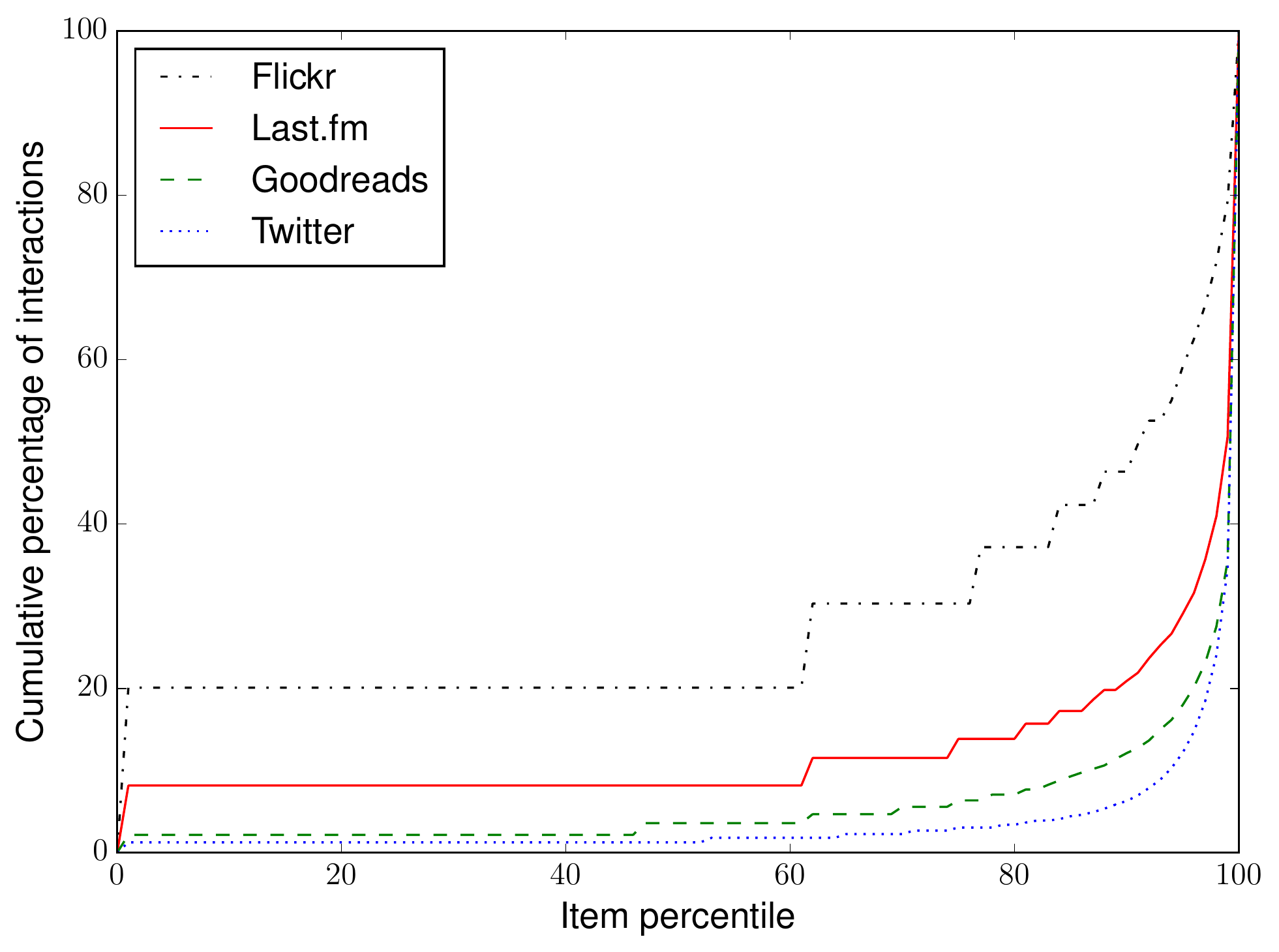}}
\caption{Cumulative percentage of adoptions by items for each dataset. Items on the x-axis are sorted by their popularity; the lines show a step pattern because multiple items may have the same number of adoptions. We observe a substantial skew in popularity. For example, the most popular 20\% of items account for 60\% of adoptions in Flickr and more than 90\% of adoptions in other datasets.}
\label{fig:percent-cumulative}
\end{figure}

Table~\ref{tab:descriptive} shows descriptive statistics about the datasets, all of which have more than 150k users and millions of items (with the exception of Twitter with 64k URLs).  Twitter has the highest mean adoptions per item ($41$), followed by Goodreads ($21$). The maximum number of adoptions for an item also varies, from more than 80k in Twitter and Goodreads to 2.7k in Flickr.  The median number of adoptions is consistent, however: at least half of the items have only 1 adoption. 
The skew in popularity distribution is better shown in Figure~\ref{fig:percent-cumulative}. The 20\% of the most popular items account for over 60\% of adoptions in Flickr and over 90\% of the adoptions in the other three websites. On Twitter, the skew is extreme: over 81\% adoptions are on 4\% of items.

\begin{figure}[t]
\centerline{\includegraphics[width=0.99\linewidth]{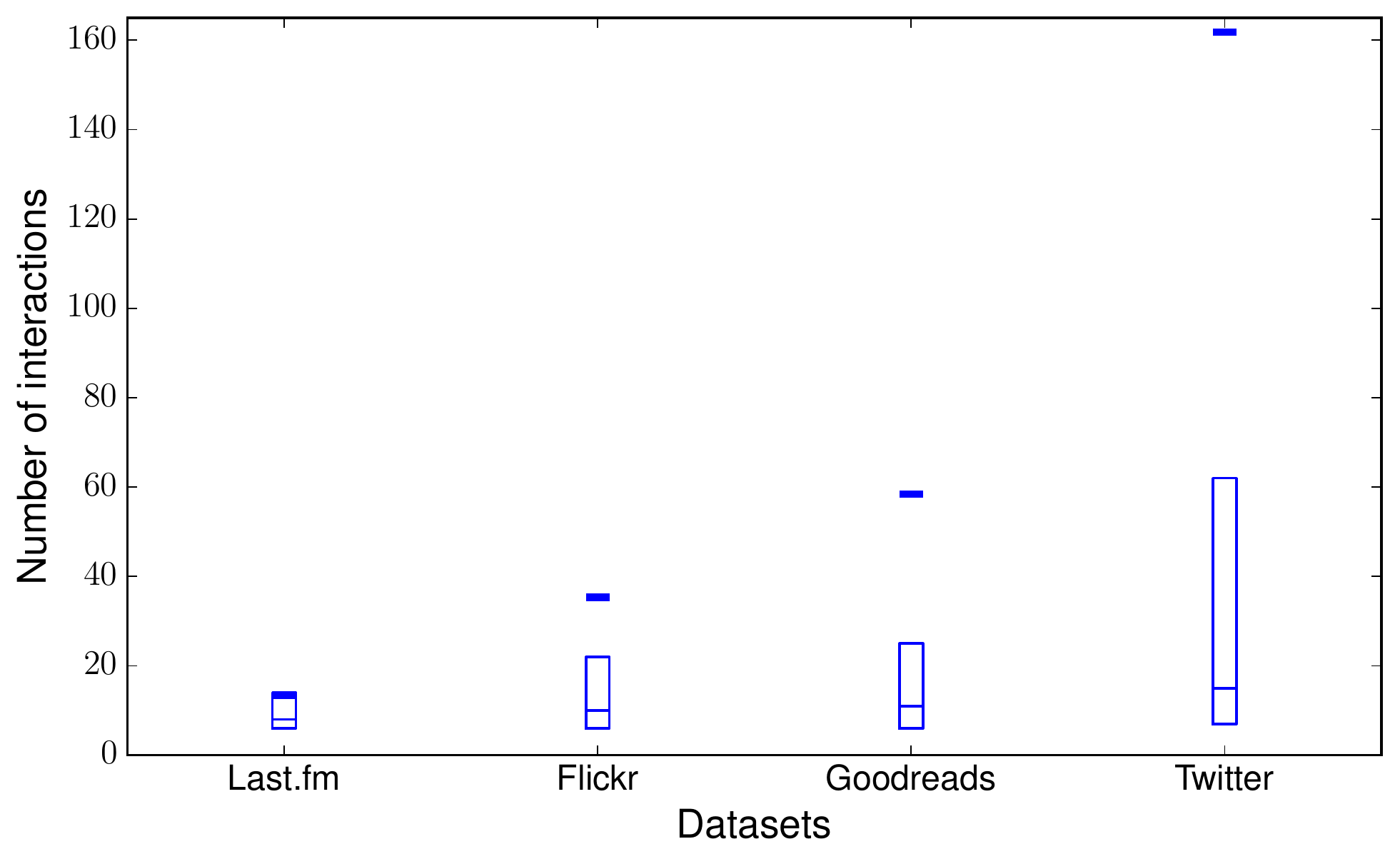}}
\caption{Boxplot showing the number of adoptions after 28 days (10 for Twitter) for items which have at least 5 adoptions. The bold partial line is the mean number of adoptions. Across datasets, most items receive less than 20 adoptions.}
\label{fig:boxplot2}
\end{figure}

\subsection{Classification methodology}
We first operationalize the Balanced Classification formulation on these datasets. As a reminder, $k$ is the number of early adoptions that we peek at for each item, and we predict which of these items will end up more popular than the median item.

We measure the final popularity at a time $T$ days after the first adoption of the item. To be consistent with prior work, we follow Cheng et al. and set $k=5$ and $T=28$ days for Last.fm, Flickr and Goodreads. Because the Twitter dataset is only three weeks long, we use a smaller $T=10$ days.
To avoid right-censoring, we include only items that had their first adoption at least $T$ days before the last recorded timestamp in each dataset.
The parameter $k$ also acts as a filter, allowing only items with at least $k$ adoptions.
Figure~\ref{fig:boxplot2} shows properties of the data thus constructed.

We classify items based on their popularity after $T$ days, labeling those above the median 1 and others as 0.  For each item, we extract features from the early adoption period, the time between the first and $k$th adoption. We use 5-fold cross validation to select the items that we train on, then use the trained model to predict final popularity of items in the test set.
Since we use median popularity as the classification threshold, the test data has a roughly equal number of items in each class, allowing us to use accuracy as a reasonable evaluation metric.
We tried several classification models using Weka \cite{hall2009weka}, including logistic regression, random forests and support vector machines.  Logistic regression models generally performed best, so we report results for those models unless otherwise specified.

\section{Balanced classification}
We start by comparing the predictive power of models using different sets of features across the four datasets on the Balanced Classification problem.

\begin{figure}[t]
\centerline{\includegraphics[width=0.99\linewidth]{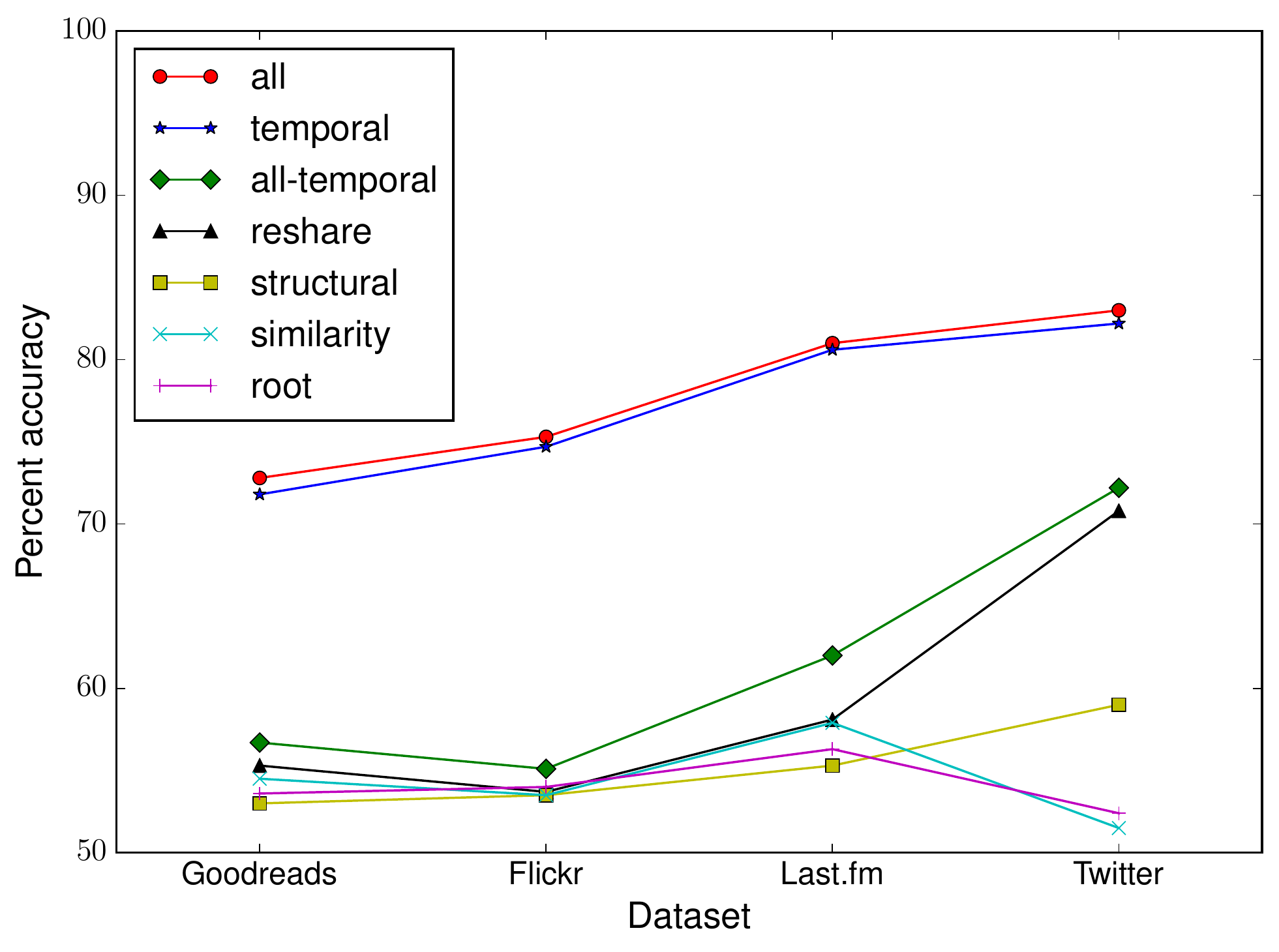}}
\caption{Accuracy for prediction models incorporating different categories of features. The y-axis starts at 50\%, the baseline for a random classifier on the balanced formulation.  On all datasets, temporal features are the most predictive, almost as accurate as using all available features.}
\label{fig:temporal-vs-all}
\end{figure}

\subsection{Temporal features dominate}

Figure~\ref{fig:temporal-vs-all} shows the prediction accuracy of the models. Similar to prior work on Facebook that used peeking \cite{cheng2014}, when using all features we are able to predict whether an item will be above the median popularity around three-fourths of the time: 73\% for Goodreads, 75\% for Flickr, 81\% for Last.fm and 83\% for Twitter.

Training models with individual feature categories shows that temporal features are by far most important. Across all four datasets, a model using only temporal features performs almost as well as the full model.  The next best feature category, resharer features, is able to predict 71\% on Twitter and less than 60\% on the other three datasets.  Even a model that uses \textit{all} non-temporal features, denoted by the ``all-temporal" line in Figure~\ref{fig:temporal-vs-all}, is not very good. For Goodreads and Flickr, this model is not much better than a random classifier.  For Last.fm and Twitter, accuracy for non-temporal features improves somewhat, but is still at least 10\% worse than when including temporal features.

Even a single temporal feature can be more predictive than models constructed from all non-temporal features.
Consider the feature $time_x$, which is the number of days for an item to receive $x$ number of adoptions.  At $x=5=k$, the feature $time_5$---time taken for an item to receive 5 adoptions---is the most predictive temporal feature for all datasets. A model based on this single feature achieves more than 70\% accuracy on all datasets and accounts for nearly 97\% of the accuracy of the full model for each dataset. While past work has highlighted the importance of temporal features as a whole \cite{szabo2010,cheng2014}, it is interesting to find that we may not even need multiple temporal features: a single measure is able to predict final popularity class label for items in all datasets.

\begin{table}[t]
\center
\begin{tabular}{lcccc} \hline
Test \textbackslash~Train & Last.fm & Flickr & Goodreads & Twitter\\ \hline
\multicolumn{5}{l}{\textbf{Using only temporal features}} \\ \hline
\ \ Last.fm & \textbf{80.6} & 80.7 & 78.0 & 80.0\\
\ \ Flickr & 73.9 & \textbf{74.7} & 70.0 & 73.9\\
\ \ Goodreads & 70.3 & 69.7 & \textbf{71.9} & 70.3\\
\ \ Twitter & 82.7 & 82.3 & 79.7 & \textbf{82.2}\\ \hline

\multicolumn{5}{l}{\textbf{Using only non-temporal features}} \\ \hline
\ \ Last.fm & \textbf{62.1} & 56.3 & 60.2 & 52.3\\
\ \ Flickr & 53.0 & \textbf{55.1} & 51.8 & 48.2\\
\ \ Goodreads & 56.0 & 52.1 & \textbf{57.1} & 50.6\\
\ \ Twitter & 45.8 & 44.1 & 56.4 & \textbf{73.4}\\ \hline
\end{tabular}
\caption{Prediction accuracy for models trained on one dataset (columns) and tested on each dataset (rows). The diagonals report accuracy on the same dataset, while other cells report accuracy when the model is trained on one dataset and tested on another.
The power of temporal features generalizes across domains: testing a model on any dataset, trained on any other dataset, loses no more than 5\% accuracy compared to testing a model on the same dataset.  For non-temporal features, prediction accuracy decreases substantially when applying models to other datasets.
}
\label{tab:train-test-temporal}
\end{table}

\section{Cross-domain prediction}
The analysis in the previous section confirms past findings about the importance of temporal features across a range of websites.
We now extend these results to show that temporal features are not only powerful, they are also general: models learnt on one item domain using temporal features are readily transferable to others. In contrast, non-temporal features do not generalize well: even the direction of their effect is not consistent across domains.  To show this, we train prediction models separately for each dataset, as before, then apply each model to every dataset.

\subsection{Temporal features generalize}
Table~\ref{tab:train-test-temporal} shows the accuracy of models trained only on temporal features from one dataset and tested on all four.  Reading across the rows shows that regardless of which social network a model was trained on, its accuracy on test data from another network remains within 5\% of the accuracy on test data from the same network.

Such consistent prediction accuracy is impressive, especially because the median time to reach 5 adoptions varies, ranging from 1 day in Flickr to 15 days for Goodreads. This suggests that there are general temporal patterns that are associated with future popularity, at least across these particular networks.

\subsection{Other features have inconsistent effects}
The story is less rosy for non-temporal features.  Table~\ref{tab:train-test-temporal} shows the cross-domain prediction accuracy for models trained on all non-temporal features (in light of their low accuracy when taken individually, we combine all non-temporal features).  Accuracies on the same dataset correspond to the ``all-temporal'' line in Figure~\ref{fig:temporal-vs-all}; they are generally low and drop further when tested on a different dataset. In particular, models trained on other websites do poorly when tested on Twitter, with the Last.fm and Flickr models performing worse than a random guesser on Twitter data.  Meanwhile, a model trained on Twitter is almost 10 percentage points worse than the Last.fm-trained model for predicting popularity on Last.fm.

Not only does  prediction accuracy drop across websites, but fitting single-feature logistic regression models for each feature shows that for 12 of the 25 features, the coefficient term  flips between being positive and negative across models fit on different datasets. Similar to the contrasting results found in prior work \cite{lerman2010,romero2013}, we find that all measures of subgraph structural features of the early adopters, namely $indegree_{sub}$, $density_{sub}$, $cc_{sub}$, $dist_{sub}$ and $sub\_in_i$ (except for $sub\_in_1$ and $sub\_in_4$), can predict either higher or lower popularity depending on the dataset. For example, a higher $density_{sub}$---number of edges in the subgraph of early adopters---is associated with higher popularity on Flickr ($\beta$ coefficient=0.04), whereas on Last.fm, a higher density is associated with lower popularity ($\beta$ coefficient=-0.09).  Features from the root, resharer and similarity categories show a similar dichotomous association with final item popularity.

\section{Gaps between prediction and understanding}
These results show that not only are non-temporal features weak predictors, the direction of their effect on popularity is inconsistent across different domains. Combining this with our observation that a single temporal heuristic is almost as good a predictor as the full model raises questions about what it is that popularity prediction models are predicting and how they contribute to our understanding of popularity.

\subsection{Temporal features drive predictability}
While our work may seem contrary to recent work that claims that early adopters and properties of their social network matter for prediction, many of their findings are consistent with our own.  Most prior work that uses peeking finds that temporal features are a key predictor \cite{tsur2012,szabo2010,pinto2013,yu2015}. Further, even though Cheng et al. conclude temporal and structural features are major predictors of cascade size, they report for predicting photos' popularity on Facebook, accuracy for temporal features alone (78\%)is nearly as good as the full model (79.5\%) \cite{cheng2014}.

By holding modeling, feature selection and problem formulation consistent, we contribute to this literature by demonstrating the magnitude and generality of the predictive power of temporal features across a range of social networks. Having multiple networks also lets us show that, unlike temporal features, using non-temporal features does not generalize well to new contexts.  These features might be useful for understanding the particulars of a given website, but it seems likely that they are capturing idiosyncrasies of that site rather than telling us something general about how items become popular in social networks.

\subsection{Is cumulative advantage the whole story?}
If non-temporal features are weakly predictive and not generalizable, and all that matters is the rate of initial adoption, then how do predictive exercises with peeking advance scientific understanding of what drives popularity?  In other words, what does it mean when one claims that popularity is predictable once we know about initial adopters?

One answer is that early, rapid adoption is a signal of intrinsic features of an item that help to determine its popularity. Items with better content command a higher initial popularity, and thus the predictive power of early temporal features is simply a reflection of content quality or interestingness to the social network in question.
Given increasing evidence from multiple domains that content features are at best weakly connected to an item's popularity \cite{salganik2006,pachet2012,martin2016}, this seems unlikely to be the whole explanation.

Another explanation is that items that get attention early are more likely to be featured in the interface, via feeds, recommendations or ads; they might also be spread through external channels could drive up the rate of early adoption. Those would be interesting questions to explore. Still, whatever be the driving reasons, these models are telling us that once items achieve initial popularity, they are much more likely to become more popular in the future.  This is simply a restatement of cumulative advantage, or the rich-get-richer phenomenon \cite{borghol2012}. 

Overall, though, we find that neither our results nor other work say much about why or how items become popular, except that items that share temporal patterns of popular items early on tend to be the ones that are more popular in the future, and that making popularity salient and ordering items by popularity can increase this effect \cite{salganik2006}.
While such predictions are practically useful for promoting content, they are not so useful for informing creation of new content or assessing its value, nor for understanding the mechanisms by which items become popular.

\section{Temporally matched balanced classification}
\label{sec:new-problems}
In this section, we give a problem formulation that lessens the importance of temporal features by conditioning on the average rate of adoption.  That is, instead of considering all items with $k$ adoptions, we consider items with $k$ adoptions within about the same amount of time.  
Given the dominance of cumulative advantage, such a formulation would be better suited for future research in understanding how items become popular, as gains in accuracy will likely shed light on attributes of early adopters, items, and networks that affect their final popularity.

\subsection{\textit{k-t} problem formulation} 

We call this formulation Temporally Matched Balanced Classification, or a \textit{k-t} formulation of the problem:
\begin{quote}
\noindent \textbf{P2: } \textit{Among items with exactly $k$ adoptions at the end of a fixed time period $t$, which ones would be higher than the median popularity at a later time $T$?}
\end{quote}

To do this, for each dataset we filtered items to those that had exactly $k$ adoptions in $t$ days.  We extracted features of these items as previously described, adding a new temporal feature for each day in $t$:
\begin{itemize}
\item $adoptions_i$: Number of adoptions on day $i$ of the early adopter period. \cite{szabo2010,tsur2012,pinto2013}
\end{itemize}

As before, we choose $k=5$ and $T=28$ days.  For each dataset, we set $t$ to be the median time it took an item to reach five adoptions: $t=15$ for Goodreads, $t=7$ for Last.fm, and $t=1$ for Flickr.  We exclude Twitter due to a lack of data when we filter for both $k$ and $t$.  We again do 5-fold cross-validation, predicting if each item would be above or below the final median popularity after $T$ days.

\begin{figure}[t]
\centerline{\includegraphics[width=0.99\linewidth]{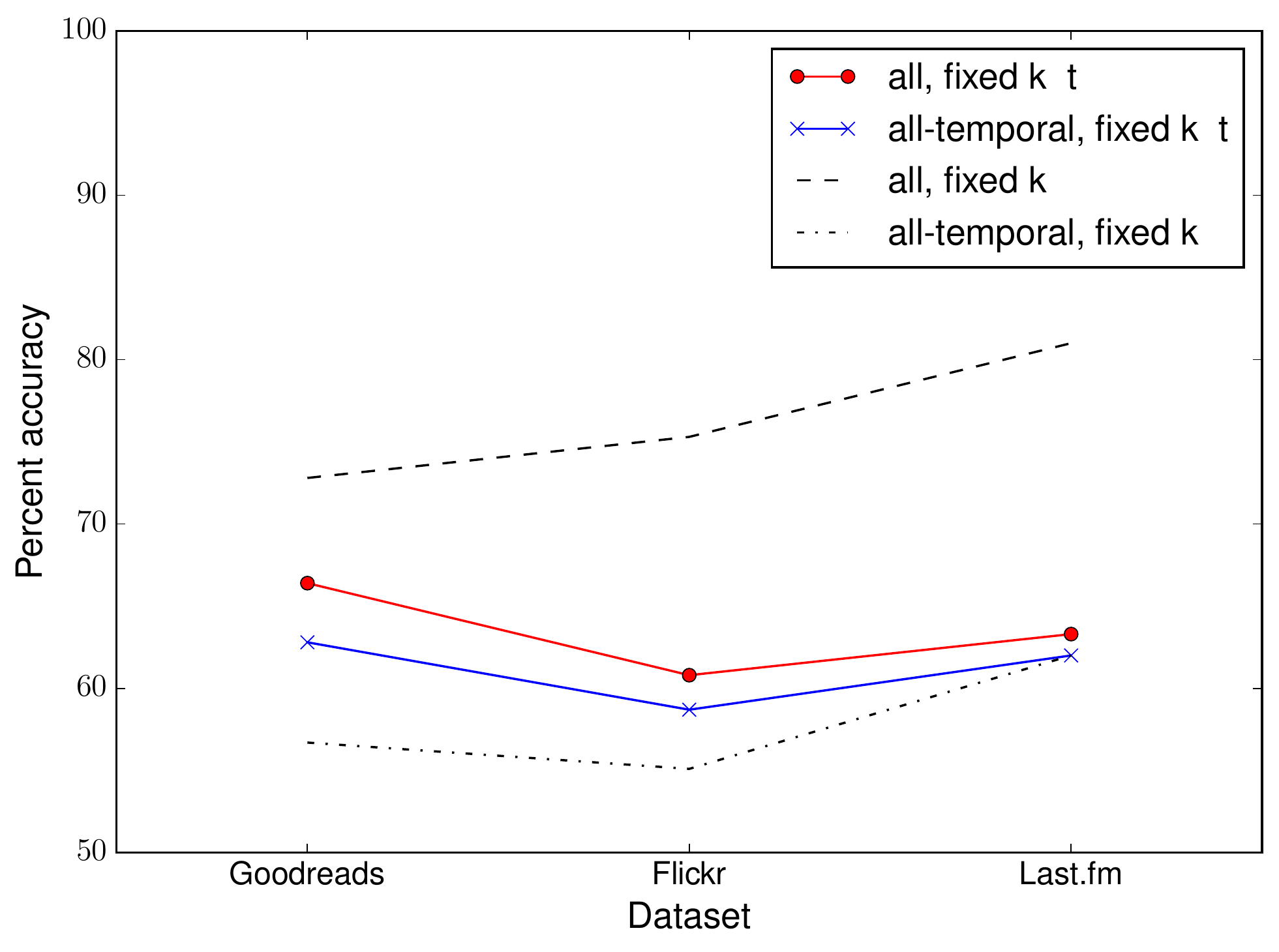}}
\caption{Percent accuracy for fixed $t$ \& $k$ using all features and non-temporal features, and for fixed $k$ with all features and non-temporal features. $k=5$, $T=28$ days for all; $t=15$ days for Goodreads, $t=1$ days for Flickr, and $t=7$ days for Last.fm.  Fixing $t$ reduces accuracy substantially compared to when $t$ is not fixed.  As expected when controlling for time, non-temporal features now provide most of the explanatory power.}

\label{fig:hard-all-temporal}
\end{figure}

Figure~\ref{fig:hard-all-temporal} shows the results.  As we hoped, non-temporal features now provide most of the explanatory power in the full model.  Further, comparing the all-temporal series with fixed $k$ and $t$ to the one with only fixed $k$ shows that the absolute accuracy of non-temporal features increases in this formulation.
This suggests that de-emphasizing temporal features in prediction might in fact improve our understanding of other features that drive popularity.

Our understanding, however, is limited: even conditioning on a single temporal feature makes for a much harder problem, with the overall prediction accuracy below 65\% for all datasets even when using all features. There is clearly much room for improvement.

\section{Discussion and Conclusion}
Using multiple problem formulations, we show that temporal features matter the most in predicting the popularity of items given data about initial adopters and our current ability to build explanatory features of those adopters and their networks.  Using datasets from a variety of social networks, we show that temporal features are not only better at predicting popularity than all other features combined, but that they readily generalize to new contexts. When we discount temporal phenomena by removing temporal features or adjusting the problem formulation, accuracy decreases substantially.

From a practical point of view, these results provide empirical support for a promising approach where only temporal features are used to predict future popularity \cite{szabo2010,zhao2015} because the drop in accuracy by casting aside non-temporal features is generally small.   Maybe creative feature engineering is not worth the effort for the Balanced Classification task.
This way of looking at the problem resonates a bit with the Netflix prize, where most of the learners that were folded into the winning model were never implemented in Netflix's actual algorithm, in part because the cost of computing and managing those learners was not worth the incremental gains \cite{amatrain2012}.

Although less valuable than temporal features, the non-temporal features examined so far do have some predictive power on their own. This might be useful when temporal information is unavailable: for example, for very new items \cite{borghol2012}, or for external observers or datasets where timestamps are unavailable \cite{cosley2010}. Encouragingly, non-temporal features increase in accuracy a little on the \textit{k-t} formulation compared to the fixed-$k$ balanced classification problem, suggesting that making time less salient might allow other factors to become more visible and modelable.  

Using \textit{k-t} models could also bend time to our advantage.  Comparing the overall performance and predictive features in models with smaller versus larger $t$ might highlight item, adopter, and network characteristics that predict faster adoption (and eventual popularity). Another way to frame this intuition is that instead of predicting eventual popularity, we should try to predict initial adoption speed.

Deeper thinking about the context of sharing might also be useful. Algorithmic and interface factors, for instance, have been shown to create cumulative advantage effects; it would be interesting to look more deeply into how system features might influence adoption behaviors. Likewise, diffusion models tend to focus attention on sharers rather than receivers of information---but those receivers' preferences, goals and attention budgets likely shape their adoption behaviors \cite{sharma2015-cscw}. Thus, consideration of audience-based features might be a way forward.  

Most generally, we encourage research in this area to go beyond the low-hanging fruit of time. For building better theories of diffusion, maximizing accuracy with temporal information may act both as a crutch that makes the problem too easy, and as a blindfold that makes it hard to examine what drives those rapid adoptions that predict eventual popularity.

\section{Acknowledgments}
This work was supported by the National Science Foundation under grants IIS 0910664 and IIS 1422484, and by a grant from Google for computational resources.

\bibliography{main}
\bibliographystyle{aaai}
\end{document}